\def\year{2022}\relax
\documentclass[letterpaper]{article} 
\usepackage{aaai22}  
\usepackage{times}  
\usepackage{helvet}  
\usepackage{courier}  
\usepackage[hyphens]{url}  
\usepackage{graphicx} 
\usepackage{natbib}  
\usepackage{caption} 
\DeclareCaptionStyle{ruled}{labelfont=normalfont,labelsep=colon,strut=off} 
\usepackage{amsfonts}
\usepackage{algorithm}
\usepackage{algorithmic}
\usepackage{graphicx}
\usepackage{textcomp}
\usepackage{algorithm}
\usepackage{multirow}
\usepackage{diagbox}
\usepackage{caption}
\usepackage{subfigure}
\usepackage{float}
\usepackage{amsthm}
\usepackage{amsmath}
\usepackage{amssymb}
\usepackage{newfloat}
\usepackage{listings}
\floatstyle{ruled}
\newfloat{listing}{tb}{lst}{}
\floatname{listing}{Listing}
\title{VGAER: Graph Neural Network Reconstruction based Community Detection}
\author{
    Chenyang Qiu,
    Zhaoci Huang,
    Wenzhe Xu,
    Huijia Li\thanks{Corresponding author.} \thanks{ This work appeared at AAAI-22: DLG-AAAI'22.}
}
\affiliations{
    Beijing University of Posts and Telecommunications \\

    Haidian, Beijing, China\\
    \{cyqiu, zc.huang, hjli\}@bupt.edu.cn \ wenzhexu@outlook.com
}

\usepackage{bibentry}

\begin{document}

\maketitle

\begin{abstract}
Community detection is a fundamental and important issue in network science, but there are only a few community detection algorithms based on graph neural networks, among which unsupervised algorithms are almost blank. By fusing the high-order modularity information with network features, this paper proposes a \textbf{V}ariational \textbf{G}raph \textbf{A}uto\textbf{E}ncoder \textbf{R}econstruction based community detection VGAER for the first time, and gives its non-probabilistic version. They do not need any prior information. We have carefully designed corresponding input features, decoder, and downstream tasks based on the community detection task and these designs are concise, natural, and perform well (NMI values under our design are improved by 59.1\% - 565.9\%). Based on a series of experiments with wide range of datasets and advanced methods, VGAER has achieved superior performance and shows strong competitiveness and potential with a simpler design. Finally, we report the results of algorithm convergence analysis and t-SNE visualization, which clearly depicted the stable performance and powerful network modularity ability of VGAER. Our codes are available at https://github.com/qcydm/VGAER.
\end{abstract}

\section{Introduction}
As one of the most important physical tools to portray the real world, network data today is gradually developing in the direction of large-scale, complexity, and modularity.  For example, the network of social platforms~\cite{wang2015community}, the protein interaction network in genetic engineering~\cite{pizzuti2014algorithms}, and the transportation network~\cite{von2009public}, etc. There are not only complex interactions between nodes, the network will also form different communities due to this interaction and node attributes. Topologically, community can be understood that internal nodes are relatively tightly connected and externally connected relatively sparse. Identifying this local structure is essential for understanding complex systems and knowledge discovery~\cite{181}.

The above task is community detection. There have been a lot of researches on community detection~\cite{Fast}~\cite{LP}~\cite{Li1}. Especially in recent years with the development of graph neural network, the graph neural network based community detection was first proposed in 2019~\cite{Chen}~\cite{Shc}, including the supervised methods: non-backtracking theories based~\cite{Chen}, Markov Random Field based~\cite{Jin} and the known unsupervised methods: complex regularization reconstruction based method GUCD(2020)~\cite{GUCD}, and negative sample contrastive learning and self-expressiveness based SEComm(2021)~\cite{SEC}. However, the prior information (such as labels) of the network community for big data systems is sometimes scarce, which also causes a great challenge to above semi-supervised methods and almost rare unsupervised methods. So it is very emergent to propose a better unsupervised method for this field.

On the other hand, after a large and comprehensive literature survey, we found that all current graph neural network community detection~\cite{Chen}~\cite{Shc}~\cite{Jin}~\cite{GUCD} and even all clustering methods like MGAE~\cite{MGAE}, ARGA~\cite{ARGA}, AGC~\cite{AGC}, SDCN~\cite{SDCN} and AGE~\cite{AGE}, etc. only focus on preserving the network structure and node features ($\mathbf A+\mathbf X$) and the clustering gain brought by different regularization methods or model design for the network embedding; fundamentally different from previous methods, we first proposed an unsupervised joint optimization method VGAER based on modularity and network structure ($\mathbf B+\mathbf A+\mathbf X$) in graph neural network based community detection which has a strict modular theoretical basis and is more suitable for the community detection, and the network structure ($\mathbf A$) is captured implicitly (by message passing phase) without being explicitly reconstructed.

We also notice that Yang proposed a nonlinear reconstruction method based on the autoencoder (denoted as DNR)~\cite{Yang} in 2016 with using this modularity theory. After 2018, on the basis of this work, a variety of autoencoder reconstruction methods that integrate different network features have been proposed~\cite{Caob}~\cite{Bha}~\cite{Caoa}. However, these methods often require additional operations and have the limited ability to capture network features, so that the $Q$ value of this DNR method is very poor when targeting a network with an unknown community structure. We empirically demonstrate this phenomenon in Section 4. To sum up, our VGAER not only has better performance comparing with these autoencoder based community detection method, but also extends to variational models that have not been covered by autoencoder based methods.

Finally, we horizontally compare the detection effects of VGAER and the advanced unsupervised GNN reconstruction based GUCD~\cite{GUCD}, which fully shows that VGAER is very competitive even compared with the most advanced and complex method, and the potential for further development (such as more complex architecture and design). And our innovative contributions and insights are as follows:
\begin{itemize}
\small
\item This paper presents a brand new community detection method based on graph variational inference, and gives a more concise non-probability version. And obtain superior performance improvements on a wide range of datasets and parameter algorithms.
\item A joint optimization framework based on modularity and network structure is proposed for the first time. And the performance improvement of VGAER benefits from nonlinear modularity reconstruction and neighborhood Laplacian smooth of our model, which can partially mitigate the extreme degeneracy problem~\cite{extre} and resolution limit~\cite{Reso} caused by single modularity maximization.
\item VGAER can not only handle community detection task, but also have a powerful generation ability, which means that VGAER can also generate the embeddings flexibly under different premises from the learned distribution. We expect in the future, VGAER can play a role in community node prediction, community embedding, personal privacy protection, etc. And we will depict a broader picture for these chances in Appendix.
\end{itemize}
\begin{figure*}
  \centering
  \includegraphics[scale=0.35]{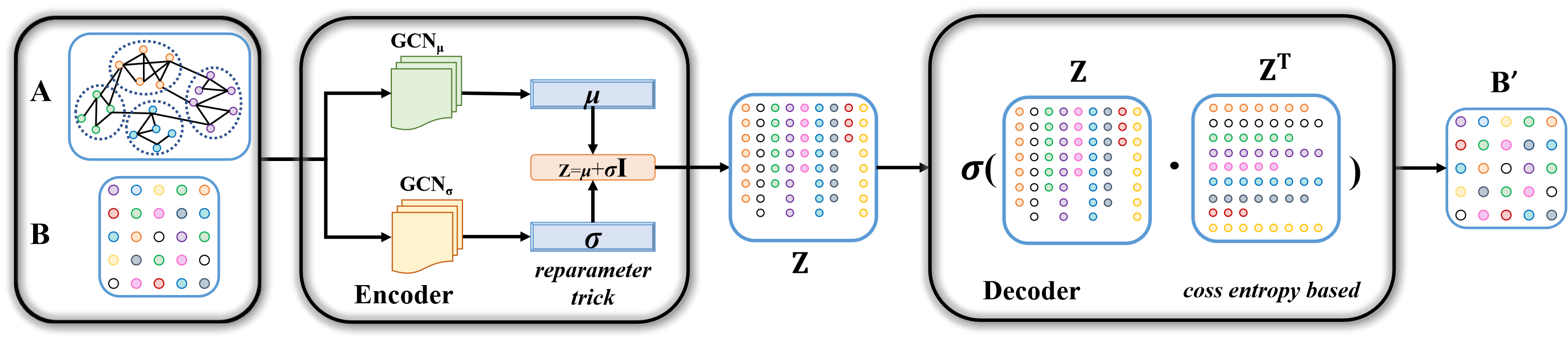}\\
  \caption{\small The architecture of the community detection model VGAER. We use the adjacency matrix $\mathbf A$ and modularity information $\mathbf B$ on the left (or concatenate the original features $\mathbf X$ of the nodes) to encode the mean vectors $\boldsymbol{\mu}$ and the standard deviation vectors $\boldsymbol{\sigma}$ of community memberships $\mathbf Z$. The modularity matrix is reconstructed by the cross entropy based decoder on the right to maximize the modularity.}
\end{figure*}

\section{Related work}
\subsection{Reconstruction and Modularity maximum}
Modularity maximum model was first introduced by Newman to maximize the modularity index $Q$ of the network~\cite{Newman}, which is defined by the following:
\begin{equation}\label{maxq}
Q=\frac{1}{2 M} \sum_{i, j}\left[\left(a_{i j}-\frac{k_{i} k_{j}}{2 M}\right) \mathcal Z \left(i,j\right)\right],
\end{equation}
where $M$ is the total number of edges of the network, $a_{i j}$ is the adjacency matrix element, a value of 1 or 0 indicates whether there is a connected edge or not, $k_{i}$ is the degree of node $i$, and $\mathcal Z$ is the association membership function of node $i$, when $i$ and $j$ belong to the same community, $\mathcal Z=1$, otherwise $\mathcal Z=0$.

Then, we simplify the Eq.(\ref{maxq}) by defining the modularity matrix $\mathbf{B}$ and introducing the node community membership vectors. Define the modularity matrix $\mathbf{B}=\left[b_{i j}\right]$ as
\begin{equation}\label{B}
b_{i j}=a_{i j}-\frac{k_{i} k_{j}}{2 m}.
\end{equation}
In this way, each node has a modularity relationship with all the other nodes, whether they have connected edges or not.

Next, we introduce a matrix $\mathbf{Z}=\left[z_{i j}\right] \in \mathbb{R}^{N \times K}$ which each row $z_i$ is the community membership vector, and $K$ is the dimension of the node community membership vector. So the Eq.~(\ref{maxq}) can be reduced as the following:
\begin{equation}\label{Q}
Q=\frac{1}{2 m} \operatorname{Tr}\left(\mathbf{Z}^{\mathrm{T}} \mathbf{B Z}\right).
\end{equation}

As a NP-hard problem, there are many different optimization ways to solve the maximization of Eq.(\ref{Q}). Here we introduce $\mathbf{Z}^{\mathrm{T}} \mathbf{Z}$ as the constant $N$ condition to relax the problem, so we obtain the following modularity optimization problem after the relaxation:
\begin{equation}
\begin{aligned}
\max Q=\max \left\{\operatorname{Tr}\left(\mathbf{Z}^{T} \mathbf{B Z}\right)\right\}\\
\text { s.t. } \operatorname{Tr}\left(\mathbf{Z}^{\mathrm{T}} \mathbf{Z}\right)=N.
\end{aligned}
\end{equation}
Based on the Rayleigh entropy, we know that the solution $\mathbf{Z}$ of the modularity degree maximization problem under relaxation conditions is the $k$ largest eigenvectors of the modularity degree matrix $\mathbf{B}$. Then, according to Eckart and Young's matrix reconstruction theorem~\cite{Young}, the equivalence of the modularity maximization and the modularity matrix reconstruction can be obtained.

\section{Methodology}
\subsection{The model}
Consider a graph $G(V,E)$, we can obtain the modularity matrix $\mathbf B$ by Eq.~\ref{B}. If $G$ have the node features $\mathbf X$, then
\begin{equation}
\mathbf{B}^{0}=\operatorname{CONCAT}(\mathbf{B},\mathbf{X}),
\end{equation}
otherwise $\mathbf{B}^{0}=\mathbf B$. We use the $\mathbf{B}^{0}$ as the input of first layer.

As a deep generative model, we firstly design a inference model for VGAER:
\begin{equation}
q(\mathbf{Z} \mid \mathbf{B}, \mathbf{A})=\prod_{i=1}^{N} q\left(\mathbf{z}_{i} \mid \mathbf{B}, \mathbf{A}\right),
\end{equation}
where $q\left(\mathbf{z}_{i} \mid \mathbf{B}, \mathbf{A}\right)$ is a variational approximation of node $i$'s true posterior distribution based on Gaussian family as:
\begin{equation}
q\left(\mathbf{z}_{i} \mid \mathbf{B}, \mathbf{A}\right)=\mathcal{N}\left(\mathbf{z}_{i} \mid \boldsymbol{\mu}_{i}, \operatorname{diag}\left(\boldsymbol{\sigma}_{i}^{2}\right)\right).
\end{equation}

Then we use two graph neural networks $\boldsymbol{\mu}=\mathrm{GCN}_{\boldsymbol{\mu}}(\mathbf{B}, \mathbf{A})$ and $\log \boldsymbol{\sigma}=\mathrm{GCN}_{\boldsymbol{\sigma}}(\mathbf{B}, \mathbf{A})$ as encoders to fit the mean vectors $\boldsymbol{\mu}$ and the standard deviation vectors $\boldsymbol{\sigma}$ for node $i$. And the encoders have the following unified form:
\begin{equation}\label{GAER-GCN}
\operatorname{GCN}\left(\mathbf{B}, \mathbf{A}\right)=\widetilde{\mathbf{A}}\operatorname{tanh}\left(\widetilde{\mathbf{A}} \mathbf{B} \mathbf{W}_{0}\right)\mathbf{W}_{1},
\end{equation}
where $\mathbf{W}_{0}$ and $\mathbf{W}_{1}$ represent the weight matrices for the first layer and second layer respectively. $\mathbf{W}_{0}$ is shared between $\mathrm{GCN}_{\boldsymbol{\sigma}}$ and $\mathrm{GCN}_{\boldsymbol{\mu}}$. $\widetilde{\mathbf{A}}=\mathbf{D}^{-\frac{1}{2}} \mathbf{(A+I) D}^{-\frac{1}{2}}$ is the symmetric renormalized adjacency matrix. And tanh is the activation function. We must point out the necessity of replacing ReLu with tanh, because the modularity matrix $\mathbf B$ contains a large number of 0 elements, and the gradient will not be updated effectively if ReLU is applied.

We can also stack multiple encoders to make VGAER fully learn the mean and standard deviation vectors of the true distribution, thereby improving its accuracy. This only requires the output of the previous encoder as the input of the next encoder.

At the deep generative stage, we specially design a cross entropy based dot product decoder to reconstruct the modularity distribution. Consider the conditional distribution of $p(B_{ij} \mid \mathbf{z}_{i} ,\mathbf{z}_{j})$, where $B_{ij}$ is a reconstructed entry, and $\mathbf z_i$ comes from a reparameterization trick. We discretize $p( B_{ij} \mid \mathbf{z}_{i} ,\mathbf{z}_{j})$ as two parts: $p( B_{ij}=b_{ij} \mid \mathbf{z}_{i} ,\mathbf{z}_{j})$ and $p( B_{ij}\neq b_{ij} \mid \mathbf{z}_{i} ,\mathbf{z}_{j})$ . At the same time, using the re-weight technique similar to VGAE~\cite{Kipf}, we re-weight these two parts with $\sigma (b_{ij})$ and (1-$\sigma (b_{ij}))$:
\begin{equation}\label{dis}
p\left(B_{i j} \mid \mathbf{z}_{i} , \mathbf{z}_{j}\right)=(\sigma\left(\mathbf{z}_{i} \mathbf{z}_{j}^{\top} \right))^{\sigma (b_{ij})} ((1-\sigma\left(\mathbf{z}_{i} \mathbf{z}_{j}^{\top}\right)))^{(1-\sigma (b_{ij}))},
\end{equation}
where $\sigma(*)=\frac{1}{1+e^{-*}}$ is a sigmoid function and the same as below. The meaning of re-weight term will be more clearer by understanding Eq.~\ref{re-weight}. And the $p(\mathbf B \mid \mathbf Z)$ is as follow:
\begin{equation}
\small
p(\mathbf{B} \mid \mathbf{Z})=\prod_{i=1}^{N} \prod_{j=1}^{N} p\left( B_{i j} \mid \mathbf{z}_{i} \mathbf{z}_{j}\right)
\end{equation}

\subsection{Optimization}
We first give the variational lower bound $\mathcal L(\boldsymbol{\phi},\boldsymbol{\theta})$ derived from the maximization objective function as follow:
\begin{equation}
\small
\begin{aligned}
&\mathbb{E}_{\mathcal B}\left[\log p_{\boldsymbol{\theta}}(\mathbf{B})\right]=\mathbb{E}_{\mathcal B}\left[\mathbb{E}_{q_{\boldsymbol{\phi}}(\mathbf{Z} \mid \mathbf{B,A})}\left[\log p_{\boldsymbol{\theta}}(\mathbf{B}, \mathbf{Z} )-\log q_{\boldsymbol{\phi}}(\mathbf{Z} \mid \mathbf{B} )\right]\right] \\
&\geq \mathcal{L}(\boldsymbol{\phi},\boldsymbol{\theta})=\mathbb{E}_{q_{\boldsymbol{\phi}}(\mathbf{Z} \mid \mathbf{B}, \mathbf{A})}[\log p_{\boldsymbol{\theta}}(\mathbf{B} \mid \mathbf{Z})]-\mathrm{KL}[q_{\boldsymbol{\phi}}(\mathbf{Z} \mid \mathbf{B}, \mathbf{A}) \| p(\mathbf{Z})],\\
\end{aligned}
\end{equation}
where $\mathcal B$ is a modularity set of $G$,$(\boldsymbol{\phi},\boldsymbol{\theta}) \in \{\mathbf W_0, \mathbf W_1, \mathbf W_2\}$ is the parameter space, and take a Gaussian prior $p(\mathbf{Z})=\prod_{i} p\left(\mathbf{z}_{\mathbf{i}}\right)=\prod_{i} \mathcal{N}\left(\mathbf{z}_{i} \mid 0, \mathbf{I}\right)$.  Then optimization task is:
\begin{equation}
\underset{\boldsymbol{\phi},\boldsymbol{\theta}}{\arg \max }\mathcal{L}(\boldsymbol{\phi},\boldsymbol{\theta}).
\end{equation}
The variational lower bound contains two terms. The first term is a reconstruction loss, and the latter term is a KL divergence that measures the similarity of two distributions. We now consider the previous specific form as follows:
\begin{equation}
\mathbb{E}_{q_{\boldsymbol{\phi}}(\mathbf{Z} \mid \mathbf{B}, \mathbf{A})}[\log p_{\boldsymbol{\theta}}(\mathbf{B} \mid \mathbf{Z})]=\mathbb{E}_{q_{\boldsymbol{\phi}}(\mathbf{Z} \mid \mathbf{B}, \mathbf{A})}[\sum_{i=1}^{N} \sum_{j=1}^{N}\log p(b_{i j} \mid \mathbf{z}_{i})].
\end{equation}
Substitute Eq.~\ref{dis} into the $\log$ term, then we can get:
\begin{equation}\label{re-weight}
\small
\log p(b_{i j}\mid \mathbf{z}_{i},\mathbf{z}_{j})=\sigma (b_{ij}) \log (\sigma\left(\mathbf{z}_{i} \mathbf{z}_{j}^{\top}\right)) + (1-\sigma (b_{ij})) \log (1-(\sigma\left(\mathbf{z}_{i} \mathbf{z}_{j}^{\top}\right))).
\end{equation}
We can also understand the above formula from the perspective of cross entropy. Eq.~(\ref{re-weight}) constructs the negative cross entropy between the true distribution $\sigma (b_{ij})$ of the entry $b_{ij}$ and the dot product reconstruction distribution $\sigma (\mathbf{z}_{i} \mathbf{z}_{j}^{\top})$. Maximizing this term is equivalent to minimizing the distance between the two distributions, that is minimizing the reconstruction loss.
\subsection{Non-probabilistic version}
We also give a non-probabilistic community detection model GAER, only one GCN as encoder is used:
\begin{equation}\label{GAER-GCN}
\mathbf Z=\operatorname{GCN}\left(\mathbf{B}, \mathbf{A}\right)=\widetilde{\mathbf{A}} \operatorname{tanh}\left(\widetilde{\mathbf{A}} \mathbf{B} \mathbf{W}_{0}\right) \mathbf{W}_{1}.
\end{equation}
And the down stream task is
\begin{equation}
\hat{\mathbf{B}}=\sigma(\mathbf Z \mathbf Z^{\top}).
\end{equation}

And this loss function performed well in experiments.

As for optimization, we suggest to use the corresponding F-norm loss for the fast community detection task, which uses the Euclidean distance between two matrices, and $\delta =\{\mathbf W_{0}, \mathbf W_{1}\}$ is the parameter space.
\begin{equation}
\begin{aligned}
\hat{\delta} &=\underset{\delta}{\arg \min } \mathcal{L}(\mathbf{B}, \hat{\mathbf{B}})=\underset{\delta}{\arg \min }\left\|\mathbf{b}^{L}\left(\mathbf{b}^{L}\right)^{T}-\mathbf{B}\right\|_{\mathrm{F}}^{2} \\
&=\underset{\delta}{\arg \min } \sum_{i=1}^{N} \sum_{j=1}^{N}\left(\hat{b}_{i j}-b_{i j}\right)^{2}.
\end{aligned}
\end{equation}

\begin{table*}[t]
\center
\caption{NMI values of 8 algorithms on 6 classic networks with known community structure}
\label{Clsssic_result}
\begin{tabular}{cccccccccc}
\hline\noalign{\smallskip}
 & VGAER &GAER &VGAECD  & VGAE  & RMOEA & GEMSEC & DNRSC & DNR \\
\noalign{\smallskip}\hline\noalign{\smallskip}
$G_1$ & \textbf{1}     &\textbf{1} &\textbf{1} & \textbf{1}     & \textbf{1} & 0.833 & \textbf{1}     & 0.692 \\
$G_2$ & \textbf{0.919} &0.877 &0.892 & 0.703 & 0.832 & 0.761 & 0.891 & 0.854 \\
$G_3$ & \textbf{0.932} & 0.904& 0.875& 0.602 & 0.749 & 0.653 & 0.864 & 0.811 \\
$G_4$ & 0.873 &\textbf{0.897} & 0.885& 0.752 & 0.734 & 0.467 & 0.627 & 0.538 \\
$G_5$ & \textbf{0.764} &0.751 & 0.758 & 0.597 & 0.372 & 0.413  & 0.337 & 0.422 \\
$G_6$ & 0.447 & 0.401 &\textbf{0.463} & 0.385 & 0.341 & 0.387 & 0.403 & 0.324 \\
\noalign{\smallskip}\hline
\end{tabular}
\end{table*}
\begin{table*}[t]
\center
\caption{$Q$ values of 8 algorithms on 6 classic networks with unknown community structure}
\label{Q_result}
\begin{tabular}{cccccccccc}
\hline\noalign{\smallskip}
 & VGAER & GAER &VGAECD & VGAE & RMOEA & GEMSEC & DNRSC & DNR  \\
\noalign{\smallskip}\hline\noalign{\smallskip}
$G_7$ & \textbf{0.5832} &0.5529& 0.5062& 0.4542 & 0.4430     & 0.4760 & 0.2854 & 0.4062      \\
$G_8$ & \textbf{0.2894} &0.2873 &0.2112 & 0.0045 & 0.0800 & 0.1900 & 0.0410 & -0.0152  \\
$G_9$ & 0.7278 &\textbf{0.8705}& 0.8438& 0.5348 & 0.7962 & 0.7580 & 0.5560 & 0.3693  \\
$G_{10}$ &\textbf{0.6436} & 0.6279& 0.6156& 0.3328 & 0.4213 & 0.3921 & 0.1237 & 0.2254  \\
$G_{11}$ & \textbf{0.8103} & 0.6991 &0.6083 & 0.3681 & 0.6827 & 0.6991 & 0.4410 & 0.1068  \\
$G_{12}$ & \textbf{0.7036}& 0.7077 & 0.6773& 0.5595 & 0.5546 & 0.4872 & 0.4549 & 0.2549   \\
\noalign{\smallskip}\hline
\end{tabular}
\end{table*}

\section{Experiment}
\subsection{Data sets and comparison algorithms}
\begin{table}[h]
\center
\caption{\small 12 real networks, where $K$ is number of communities, $N$ is number of nodes, and $M$ is number of edges, C represents whether community is known(K) or not(C).}
\label{Clsssic}
\begin{tabular}{cccccc}
\hline\noalign{\smallskip}
Symbol & Dataset & K & N & M & C \\
\noalign{\smallskip}\hline\noalign{\smallskip}
$G_1$ & Karate & 2 & 34 & 78 & K \\
$G_2$ & Dolphins & 2 & 62 & 159 & K \\
$G_3$ & Friendship & 6 & 68 & 220 & K \\
$G_4$ & Football & 12 & 115 & 613 & K \\
$G_5$ & Polblogs & 2 & 1490 & 16718 & K \\
$G_6$ & Cora & 7 & 2708 & 5429 & K \\
$G_7$ & Les Miserables & - & 77 & 254 & U \\
$G_8$ & Adjnoun & - & 112 & 425 & U \\
$G_9$ & Netscience & - & 1589 & 2742 & U \\
$G_{10}$ & PPI & - & 2361 & 7182 & U \\
$G_{11}$ & Power Grid & - & 4941 & 6594 & U \\
$G_{12}$ & Lastfm\_asia & - & 7624 & 27806 & U \\
\noalign{\smallskip}\hline
\end{tabular}
\end{table}

We selected 12 real-world network datasets, as shown in Table~\ref{Clsssic}, which separately have their characteristics, thus giving the algorithms different challenges to more comprehensively evaluate the effects of each algorithm, and half of them have unknown community structure. 

As for comparison algorithms, we choose other 6 kinds of methods, which are representative and can be divided into two categories: one is the advanced non-deep learning community detection method: they are graph embedded based GEMSEC~\cite{GEMSEC}, multi-objective evolutionary based algorithm RMOEA~\cite{RMOEA}; the other is deep learning-based nonlinear method: autoencoder reconstruction based DNR~\cite{Yang}, autoencoder method with incorporating network structure and nodes contents DNRSC~\cite{DANMF} and graph neural network based method VGAE~\cite{Kipf} and VGAECD~\cite{VGAECD}.

\subsection{Evaluating on networks}
Table~\ref{Clsssic_result} shows the NMI results on known-community-structure networks, VGAER and GAER achieves the best standardized mutual information value NMI on most real networks. This shows the ability of VGAER to correctly divide nodes with known community structures. 

Table~\ref{Q_result} shows the $Q$ value results for the 6 networks with unknown community structure, VGAER and GAER achieved the best $Q$ value among the 6 networks. This shows that VGAER can accurately find the tight modular structure (community) in the network. On the other hand, the semi-supervised method will completely fail in this kind of network, while VGAER can find close communities and other interesting structures.
\begin{figure}[h]
  \centering
\  \includegraphics[scale=0.38]{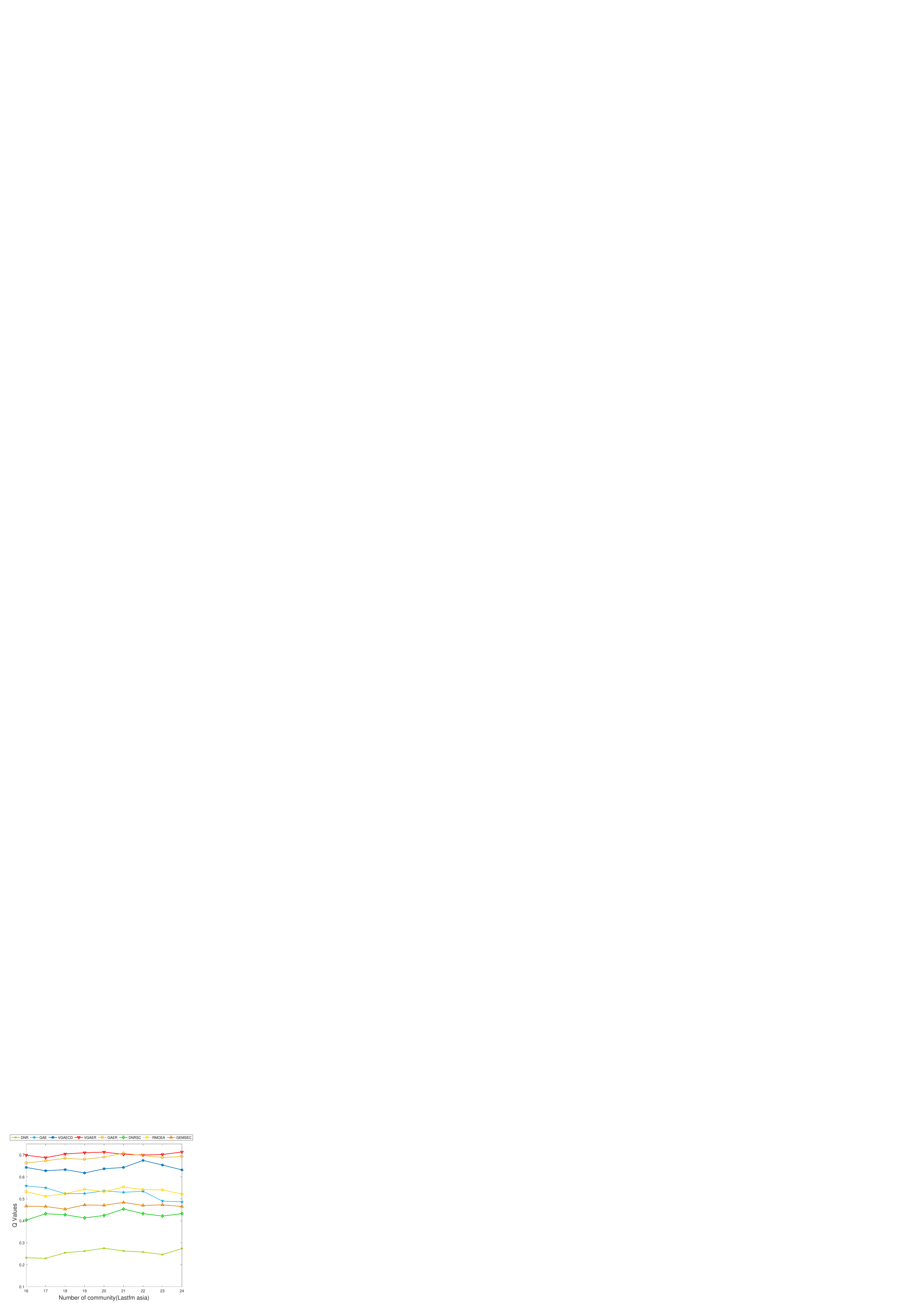}\\
  \caption{\small The $Q$ values on Lastfm\_asia with different community numbers.}\label{ASIA}
\end{figure}

\begin{figure*}[t]
  \centering
  \includegraphics[scale=0.3]{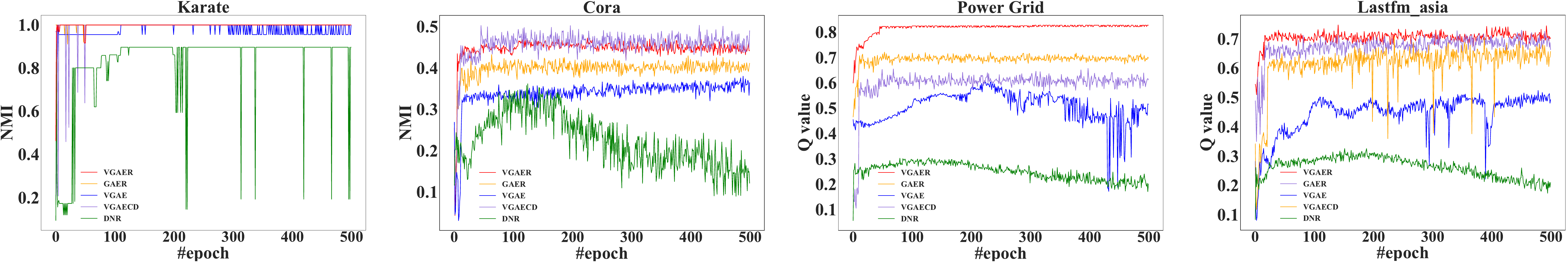}\\
  \caption{The Q and NMI values of algorithms in different epoch on four networks.}\label{convergence}
\end{figure*}
\begin{figure*}
  \centering
  \includegraphics[scale=0.29]{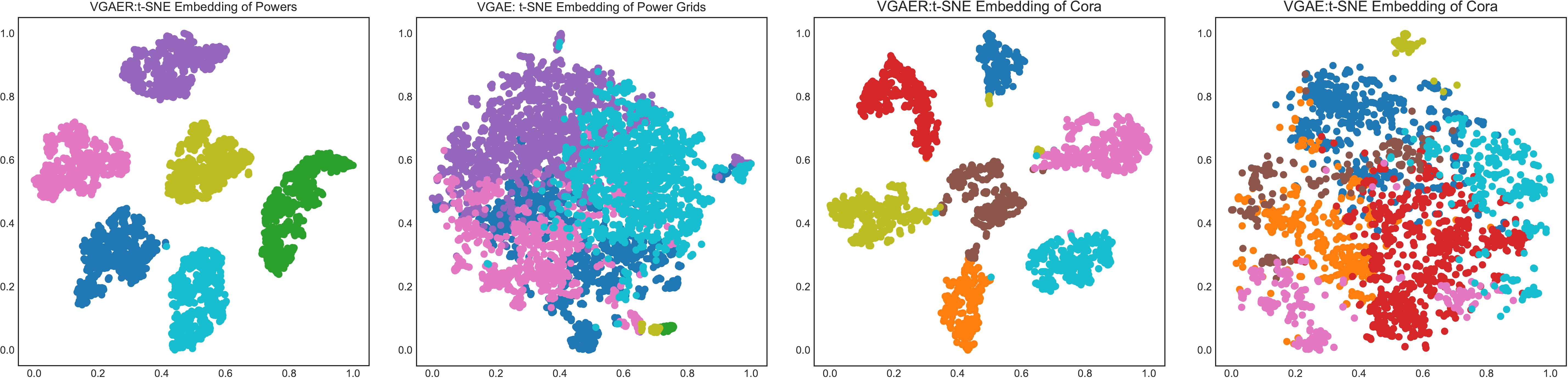}\\
  \caption{t-SNE visualization on Power Grids and Cora}\label{tsne}
\end{figure*}
At the same time, we horizontally compare VGAE and autoencoder family (DNR,DNRSC), which as part of the inspiration for the VGAER algorithm. First of all, the $Q$ values of autoencoder family are very poor in various unknown-community networks. On the other hand, VGAE performs slightly better in some networks, which suggests that in community detection tasks, the use of network structure still has great utility. But the results of VGAE are still far worse than VGAER: as for the $G_8$ network, the $Q$ value of VGAE is almost 0, the community structure cannot be detected; and excluding $G_8$, VGAER has an improvement of 25.76\%-120.13\% over VGAE. which fully proves the effectiveness of the GAER model in community detection, rather than a trivial extension of the VGAE model. These comparisons can also be obtained clearly from Fig.~\ref{convergence}.

At last, we focused on analyzing a real social network $G_{12}$. It has always been an interesting topic to find close communities from social networks to guide marketing, relationship discovery, etc. We set a optimal community range, and a series of experiments were conducted. We report the results in the fig.\ref{ASIA}, VGAER achieves excellent results near 0.7 under each number setting ($Q$ value on the real network is generally 0.3 to 0.7 caused by sparseness of the network edge).

\subsection{Algorithm analysis}
We first report the Q or NMI values of each epoch on four networks to show the convergence of VGAER in Fig.~\ref{convergence}. It can be seen clearly from the Fig.~\ref{convergence} that the Q or NMI results of VGAER after convergence is high and stable, and the initial Q or NMI values of VGAER are relatively high. 

Then we report t-SNE~\cite{SNE} visualization on Power Grids and Cora. It can be seen from Fig.~\ref{tsne} that the embeddings of VGAE are mixed together and basically indistinguishable, which may be related to the reconstruction of the adjacency matrix $\mathbf A$. While the embedding of VGAER forms a highly modular cluster structure, which is strongly distinguishable. The visualization results not only show the powerful modularity ability of VGAER, but also reveal the rationality of reconstructing the modularity matrix $\mathbf B$. This can be understood from downstream tasks, because the original task of VGAE is link prediction, so the reconstruction of relationships between nodes is helpful. However the community detection task pays more attention to the higher-order relationship, and it may not be the most wise choice to continue to reconstruct the adjacency matrix (including VGAECD).
\begin{table}[b]
\center
\caption{\small The NMI values on three popular datasets between VGAER and the most advanced method GUCD.}
\label{Simple}
\begin{tabular}{cccc}
\hline\noalign{\smallskip}
Dataset & GUCD & VGAER & Constract Ratio  \\
\noalign{\smallskip}\hline\noalign{\smallskip}
Cora & 0.3233 & \textbf{0.4468} & $\uparrow$ 38.20\%  \\
Citeseer & \textbf{0.2743} & 0.2169 & $\downarrow$ 20.93\%  \\
Pubmed & 0.2698 & \textbf{0.2729} & $\uparrow$ 1.14\%  \\
\noalign{\smallskip}\hline
\end{tabular}
\end{table}
\subsection{Further discussion}
As a joint optimization method coming from modularity maximization theory, VGAER's high Q values and powerful network modularity ability (visualization) on the unknown-community networks are impressive; at the same time, the performance of VGAER is also good and stable on networks with ground truth. But we also admit that simple modularity reconstruction cannot yet achieve the best NMI performance on some networks with ground truth, which \textbf{leaves a lot of room for improvement in VGAER}. On the other hand, despite this, in the existing reconstruction-based unsupervised graph neural network community detection method, VGAER is still a simple but effective method which shows a strong competitiveness. As shown in the Table.~\ref{Simple}, where GUCD~\cite{GUCD} (\emph{IJCAI}) may be now the only unsupervised reconstruction based community detection method except VGAER, and entire framework of GUCD is very complicated. But VGAER still has two of networks surpassed GUCD.

\section{Conclusion}
This paper provide two options for unsupervised community detection based on graph neural networks: one is a graph variational inference based VGAER, which is not limited to handle community detection, the research on VGAER under different conditions generating embeddings will be an interesting and useful topic; the other is a faster and non-probabilistic GAER. And the NMI and Q results of VGAER and GAER on real world networks are impressively high and stable. We will further study the more refined division strategy for ground-truth networks on VGAER and the semi-supervised framework. 
\section*{Acknowledgment}
We are grateful to the anonymous reviewers for their professional and valuable suggestions for improving the manuscript. And this work was partially supported by National Natural Science Foundation of China under Grant 71871233, and Fundamental Research Funds for the Central Universities of China under Grant 2020XD-A01-2.

\section{Reference}
\nobibliography*
\bibentry{wang2015community}\\
\bibentry{pizzuti2014algorithms}\\
\bibentry{von2009public}\\
\bibentry{181}\\
\bibentry{Li1}\\
\bibentry{Fast}\\
\bibentry{LP}\\
\bibentry{Chen}\\
\bibentry{Shc}\\
\bibentry{Jin}\\
\bibentry{GUCD}\\
\bibentry{SEC}\\
\bibentry{MGAE}\\
\bibentry{ARGA}\\
\bibentry{AGC}\\
\bibentry{SDCN}\\
\bibentry{AGE}\\
\bibentry{Yang}\\
\bibentry{Caob}\\
\bibentry{Bha}\\
\bibentry{Caoa}\\
\bibentry{extre}\\
\bibentry{Reso}\\
\bibentry{Newman}\\
\bibentry{Young}\\
\bibentry{Kipf}\\
\bibentry{GEMSEC}\\
\bibentry{RMOEA}\\
\bibentry{DANMF}\\
\bibentry{SAGE}\\
\bibentry{VGAECD}\\
\bibentry{SNE}\\
\bibentry{GAT}\\
\bibentry{PAM}\\
\bibentry{GIN}\\
\bibentry{MAE}\\
\bibentry{CM}\\
\bibentry{CM2}\\
\bibentry{privacy1}\\
\bibentry{privacy2}
\nobibliography{aaai22}

\clearpage

\section{Appendix}
\appendix

\section{Different encoding modules for VGAER}
The essence of VGAER method is to obtain the low-order embedding probability of modularity (and feature connection) through the encoder based on graph neural network, and use this low-order embedding to reconstruct the distribution of modularity matrix. Therefore, VGAER allows us to use any GNN model or plug and play module in the encoding stage, such as: \textbf{GraphSAGE}~\cite{SAGE}, which realizes inductive learning through neighborhood sampling encoding and greatly reduces the complexity of the algorithm. We can see the extension of VGAER on GraphSAGE in the next section for the scalability; \textbf{GAT}~\cite{GAT}, which enables VGAER to capture the weights between different node modules; \textbf{PAM}~\cite{PAM}, which is actually a plug and play module with linear complexity, and faster incremental community detection can be achieved on the basis of GraphSAGE inductive learning; \textbf{GIN}~\cite{GIN}, which introduce a $\epsilon$ to realize the injectivity of aggregation, thereby achieving a more powerful aggregation function.
\section{Does the cross entropy based decoder really work?}
Then we demonstrate the effectiveness of the cross entropy based decoder through a set of simple experiments. As shown in the following Table~\ref{Work}, VGAER uses the cross entropy based decoder, and VGAER(dot) uses the 0-1 distributed dot product based decoder (following the design of VGAE).
\begin{table}[h]
\center
\caption{The NMI and Q values of networks on two different designed decoders}
\label{Work}
\small
\begin{tabular}{ccccc}
\hline\noalign{\smallskip}
\multirow{2}{*}{Dataset} & \multicolumn{2}{c}{VGAER} & \multicolumn{2}{c}{VGAER(dot)}   \\
\cline{2-5}
 &NMI &Q &NMI &Q \\
\noalign{\smallskip}\hline\noalign{\smallskip}
Cora&0.447($\uparrow$ 59.1\%) &0.657($\uparrow$ 3.2\%) &0.281 & 0.636 \\
Citeseer&0.227($\uparrow$ 187.3\%) &0.610($\uparrow$ 7.8\%) &0.079 & 0.566\\
Pubmed&0.273($\uparrow$ 565.9\%) &0.558($\uparrow$ 66.1\%) &0.041 & 0.336 \\
\noalign{\smallskip}\hline
\end{tabular}
\end{table}

The results in the Table~\ref{Work} are interesting and non-trivial: firstly, the experiments on multiple datasets fully depict the effectiveness of the cross entropy based decoder we designed, and its NMI and Q on all datasets are significantly higher than those of the original VGAE's dot decoder. Secondly, the influence of the cross entropy based decoder on NMI (59.1\%-565.9\%) seems to be significantly greater than its influence on the Q values (3.2\%-66.1\%), which means that when the cross entropy based decoder is not used, most of the networks with ground truth will have a totally wrong division (NMI close to 0).

\section{Scalability}
As a reconstruction method, the time complexity of the original VGAER is $O(MN)$ and the space complexity is $O(N^2)$. Therefore, we analyze the source of complexity in detail and introduce some techniques to enhance the scalability of VGAER.

\textbf{Complexity analyze:}
\begin{itemize}
\item Input feature: the dimension of input feature ($O(N)$) is the main part of the time complexity of the algorithm. In the future work, how to compress the dimension of the input features as much as possible under the condition of ensuring accuracy will be of great practical significance.
\item Matrix product: Neighbor smooth sampling is an important step to obtain low rank community attribution coding. However, the frequency domain method represented by GCN will lead to high complexity by matrix multiplication. An effective measure is the sampling strategy of fixed number of neighbors ($k$) represented by GraphSAGE.
\item Matrix reconstruction: Matrix reconstruction is the main part of spatial complexity. Reducing the spatial complexity of VGAER is a issue. Theoretically, even if we adopt a mini-batch training strategy, we have to save other reconstruction elements that have not been updated for gradient update. At the same time, we have also noticed the recent work in computer vision~\cite{MAE}. Image reconstruction is realized through asymmetric encoder and decoder and good performance is guaranteed. This asymmetric design enables MAE~\cite{MAE} to be extended to large-scale images. Therefore, it is interesting to study the feasibility of introducing asymmetric encoders and the methods to ensure efficiency and accuracy in community detection.
\end{itemize}

We recommend the following techniques to enhance the scalability of VGAER. The time complexity of VGAER using the following methods will be reduced to $O(kN+pNF)$, where $k$ is the number of neighborhood samples, $p$ is the number of each batch, and $F$ is the low dimensional attribution coding dimension.

\textbf{Scalability techniques:}
\begin{itemize}
\item $k$ Neighbors sampling
\item Mini-batch training
\item Stochastic gradient descent
\end{itemize}

Now, we show the $k$ neighbors sampling technique. We decompose one-step encoding into two-stage encoding: \textbf{Neighborhood Sharing} and \textbf{Membership Encoding}, through a limited number of neighbor sampling $k$; we will obtain condensed and high-quality neighborhood information.

The Neighborhood Sharing stage is as follow:
\begin{equation}\label{NS}
\mathcal P_{n(v)}^{l} = \mathcal NS\left(\left\{\mathcal P_{u}^{l-1}, \forall u \in \mathcal {N}(v)\right\}\right).
\end{equation}
where $\mathcal P= \{\boldsymbol{\sigma}, \boldsymbol{\mu}\}$ is the set of variance and mean. $\mathcal NS$ is a Neighborhood Sharing operator. We use the MEAN as $\mathcal NS$, and $\mathcal {N}(v)$ represents a finite neighbors set of node $v$, and the number is usually $k$. The other operators are recommended as LSTM, Pool~\cite{SAGE} and GAT~\cite{GAT} ect.

And the Membership Encoding stage is as follow:

When $l=1$:
\begin{equation}\label{ME}
\mathcal P_{v}^{1} = \sigma\left(\operatorname{CONCAT}\left(\mathcal P_{v}^{0}, \mathcal P_{\mathcal {N}(v)}^{1}\right) \mathbf{\cdot W}_{0}\right).
\end{equation}

When $l=2$:
\begin{equation}\label{ME1}
\mathcal P_{v}^{2} = \operatorname{CONCAT}\left(\mathcal P_{v}^{1}, \mathcal P_{\mathcal {N}(v)}^{2}\right) \mathbf{\cdot W}_{i},
\end{equation}
where $i=1$ or 2 represents the second-layer weight matrix of variance and mean respectively.

The pseudo-code of the two-stage VGAER framework is showed in Algorithm~\ref{alg1}.
\begin{algorithm}
	\renewcommand{\algorithmicrequire}{\textbf{Input:}}
	\renewcommand{\algorithmicensure}{\textbf{Output:}}
	\caption{Framework of the two-stage VGAER}
	\label{alg1}
	\begin{algorithmic}[1]
        \REQUIRE  Graph $G(V, E)$; node features $\left\{\mathrm{x}_{v}, \forall v \in V\right\} ;$ layer depth $L$; weight matrices $\mathbf{W}_{i}, \forall i \in\{0,1,2\}$; non-linearity $\sigma$; neighborhood node set: $\mathcal {N}(v)$.
        \ENSURE  Community membership vectors ${z}_{v}$ for all $\forall v \in V$
		\FOR  {$l$ =1,2}
		\FOR {$\forall v \in V$}
		\STATE \textbf{Neighborhood Sharing}: Sample the neighborhood codes of node $v$, and obtain neighbor information $\mathcal P_{\mathcal {N}(v)}^{l}$ by Eq.(\ref{NS})
		\STATE \textbf{Membership Encoding}: Merge of $v$ and its neighborhood, and obtain low-rank membership encoding $\mathcal P_{v}^{l}$ by Eq.(\ref{ME}) and Eq.(\ref{ME1})
		\ENDFOR
		\ENDFOR
		\STATE   $\mathbf{z}_{v} \leftarrow {\boldsymbol{\mu}_v}+{\boldsymbol{\sigma}_v} \circ \boldsymbol{\epsilon}_{v}, \forall v \in V$, $\{\boldsymbol{\mu}_v, \boldsymbol{\sigma}_v\} \in \mathcal P_{v}$
	\end{algorithmic}
\end{algorithm}

Based on Algorithm 1, the more details of VGAER's mini-batch stochastic gradient training can be found in~\cite{SAGE}.

In short, with the use of these scalability technologies, the time complexity of each forward propagation of VGAER is $O(N)$, which is linear with the network nodes. However, if we want to further enhance the speed and scalability of VGAER (for example, the forward propagation time complexity reaches the sublinear or constant order $O(1)$), we must further consider the feasibility and method of compressing the input features.

At last, this two-stage inductive learning strategy allows us to implement incremental community detection for new nodes in the network.
\section{Broader Impact}
\subsection{Community node prediction}
 As a high-level structure, the community is different from the label, which also makes it difficult to directly train the community node prediction. Therefore, for new nodes in the network, in addition to retraining the entire graph (which is almost impossible to achieve on large-scale networks), incremental clustering methods such as incremental K-Means are used for incremental community detection. But VGAER's generative ability gives us new inspiration. Can we improve the prediction accuracy of community nodes by designing a better measurement method between the generation distributions instead of naive low-dimensional representation similarity distance?

\subsection{Community embedding}
\begin{figure}[h]
  \centering
  \includegraphics[scale=0.3]{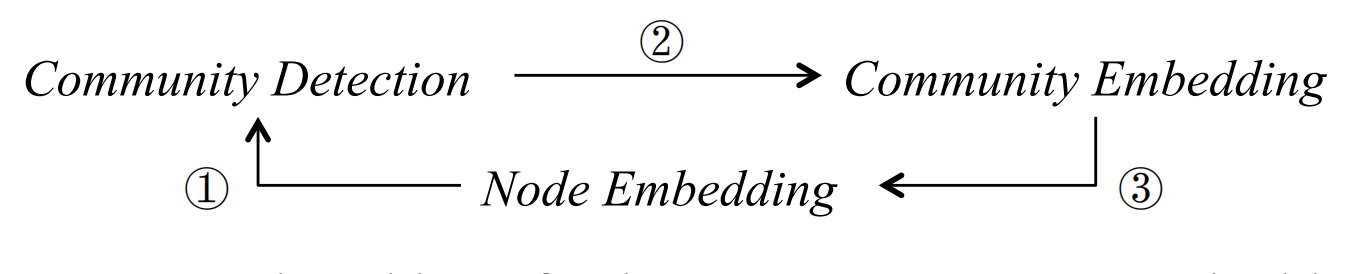}\\
  \caption{Closing loop for community embedding, community detection and node embedding (from~\cite{CM}).}\label{loop}
\end{figure}

Different from node embedding and graph embedding, community embedding aims to obtain the embedding representation of the community structure in the network, which is the high-level information of the network between node-level and graph-level embedding. The obtained community embedding will further strengthen the community detection effect, which also helps to improve the node embedding effect, can be represented by a closed loop. However, most of the existing community embedding methods are to first obtain the node embedding, and then use the mixed multi-Gaussian distribution to fit the community embedding distribution~\cite{CM}~\cite{CM2}, and VGAER can directly obtain the node-related community distribution. Therefore, how to design the more direct and concise community embedding on the basis of VGAER will further improve the Fig.~\ref{loop}'s closed-loop effect.

\subsection{Personal privacy protection}
As a network cluster structure, communities are critical in finance, interpersonal communication, and merchandise sales. However, explicit node embedding often leads to privacy leakage and personal information tracking. For example, in a social network, explicit node embedding is often contains condensed personal information, and through simple similarity calculations with other nodes, we can learn about personal potential social relationships~\cite{privacy1}. We have noticed that some of the current privacy protection strategies are implemented through generative models~\cite{privacy1}~\cite{privacy2}. Therefore, how to incorporating VGAER's generation ability into privacy protection area to further improve user's privacy and security in community detection will have great practical significance.

\end{document}